\documentclass{article}
\usepackage{spconf,amsmath,graphicx}
\usepackage{color}
\usepackage{url}

\def\START{{\tt <START> }}
\def\END{{\tt <END> }}
\def\PAD{{\tt <PAD> }}
\def\UNK{{\tt <UNK> }}
\def\CLS{{\tt [CLS] }}

\title{Automatic Audio Captioning using Attention Weighted Event based Embeddings}
\name{Swapnil Bhosale, Rupayan Chakraborty, Sunil Kumar Kopparapu}
\address{TCS Research, Tata Consultancy Services Limited, India.}
\begin{document}
\ninept
\maketitle
\begin{abstract}
Automatic Audio Captioning (AAC) refers to the task of translating audio into a natural language that describes the audio events, source of the events and their relationships. The limited samples in AAC datasets at present, has set up a trend to incorporate transfer learning with Audio Event Detection (AED) as a parent task. Towards this direction, in this paper, we propose an encoder-decoder architecture with light-weight (i.e. with lesser learnable parameters) 
Bi-LSTM recurrent layers for AAC and compare the performance of two state-of-the-art pre-trained AED models as embedding extractors. Our results show that an efficient AED based embedding extractor combined with temporal attention and augmentation techniques is able to surpass existing literature with computationally intensive architectures. % with higher learnable parameters. 
Further, we provide evidence  of the ability of the non-uniform attention weighted encoding generated as a part of our model to facilitate the decoder glance over specific sections of the audio while generating each token.
\end{abstract}
\section{Introduction}
\vspace{-0.2cm}
\label{sec:intro}
Automatic Audio Captioning (AAC) is an inter-modal translation task, where the objective is to generate a textual description for a corresponding input audio signal \cite{drossos2017automated}. Audio captioning is a critical step towards machine intelligence and many applications in daily scenarios,
such as audio retrieval \cite{oncescu2021audio}, scene understanding \cite{wu2019enhancing}\cite{lu2015context}, applications for the hearing impaired patients \cite{hong2010dynamic}, detailed audio surveillance etc.
Unlike an Automatic Speech Recognition (ASR) task, the output is a description rather than a transcription of the contents within the audio sample. An ASR task recognizes the
linguistic content (text) of the human spoken utterance; any background audio events are considered
to be "noise" and removed during pre-processing. A precursor to the AAC task is the Audio Event Detection (AED) \cite{portelo2009non}\cite{babaee2017overview} problem, with emphasis on categorizing a sound into a set of pre-defined audio event labels. An audio caption includes but is not limited to, identifying the presence
of multiple audio events ("dog bark", "gun shot" etc.), acoustic scenes ("in a crowded place", "amidst heavy rain" etc.), the spatio-temporal relationships of event source, ability to distinguish foreground audio events, from background audio events ("kids playing, while birds chirping in the background"), and physical properties based on the interaction of the source objects with the environment (" door creaks as it slowly revolves back and forth") \cite{drossos2020clotho}\cite{kim2019audiocaps}.%AAC in addition to detection of unique audio events, emphasizes on the states, actions and interaction within different audio event's source, or between the source and the environment.

%\section{Related Work}
%\label{sec:related_work}
Caption generation is an integral part of scene understanding which involves perceiving the relationships between actors and entities. It has primarily been modeled as generating natural language descriptions using image or video cues \cite{chen2019temporal}. However, audio based captioning was recently introduced in \cite{drossos2017automated}, as a task of generating meaningful textual descriptions for audio clips. In the remaining section, we provide a brief overview of existing literature for caption generation with auto-regressive modeling i.e., making predictions based on the outputs at previous time steps, followed by recent developments in transfer learning based approaches for AAC.

Autoregressive encoder-decoder architectures \cite{gregor2014deep} generate a single token for every forward pass through the decoder, and undergo subsequent decoding steps to obtain the entire caption. The raw features are first fed as input into the encoder which computes a bottleneck encoding. The input to the decoder is the encoded feature, and the token generated at previous time-step. The decoding is continued unless the model outputs an \END token or reaches a predefined maximum length for caption. Recent trends \cite{xu2021investigating}\cite{koizumi2020audio}\cite{mei2021encoder}, primarily owing to the expensive task of generating captions, have been at making use of information for other similar tasks within audio understanding, such as audio tagging \cite{fonseca2018general}, or acoustic scene classification \cite{barchiesi2015acoustic}.

In \cite{koizumi2020transformer} a transformer based audio captioning model, with keyword estimation (related to the audio event) as an auxiliary task
was proposed. During training the keywords (i.e. targets for auxiliary tasks) are chosen using heuristics from the training audio caption, while during testing, the model estimates keywords which are referred during caption generation. \cite{ozkaya2021audio} on the other hand combines the encoder-decoder architecture with semantic information extracted from the subject-verb embeddings from audio captions during the training. Their model yields substantially improved captioning performance compared to when using raw audio features, and produces captions with similar concepts and acoustically similar audio events. However, the smaller size of the available AAC datasets, has made the use of complex architectures for training AAC tasks from scratch difficult. Inspired by the rising trend in favor of using transfer learning within the image captioning literature \cite{hossain2019comprehensive}\cite{xu2015show}, recent works within AAC have also proposed use of pre-trained AED based models for transfer learning. Most recently, \cite{xu2021investigating} explored transfer learning for AAC, by considering two different pre-trained models on the AED tasks. The authors hypothesize that the use of audio tagging and acoustic scene classification tasks aids in learning the local information related to individual sound events and the global information capturing the acoustic scenery, respectively. Their models trained separately on each of the two tasks weigh the importance of local information in the context of AAC higher than the global information.
 
The main contributions of this work are, (1) we propose a light weight encoder-decoder model with attention using only recurrent layers which performs competitively with existing literature with complex architectures% having much larger set of trainable parameters
, (2) we compare two different pre-trained AED models 
(YAMNet \cite{yamnet2019} and the more recent Audio Spectrogram Transformer (AST) \cite{gong2021ast})
for transfer learning on the AAC task,
%(a recently proposed state-of-the-art AED model). 
To the best of our knowledge, this is the first attempt of adapting the AST architecture for another downstream translation task,
and
%The performance of the AAC task is heavily dependent on the performance of the AED model used for transfer learning, 
(3) we provide an intuitive view of the attention weights along with the step-by-step decoding process and testify the ability of our model to localize specific events within the entire duration of the audio and match it with specific keywords generated as a part of the output text sequence. Interestingly, the visualizations provide an holistic view  specifically, in cases pertaining to incorrectly generated captions as a result of the model's inability to identify (miss or mis-classify) audio events within the input audio.

%The rest of the paper is organized as follows, in Section \ref{sec:system_design}, we explain in detail the system design and our proposed approach. We provide the training and evaluation details and solidify the intuitions for temporal attention in context of caption generation in Section \ref{sec:experiments}. We finally conclude in Section \ref{sec:conclusion}.

\section{System Design}
\begin{figure}[t]
	\centering
	%\centerline{
	\includegraphics[width=0.48\textwidth]{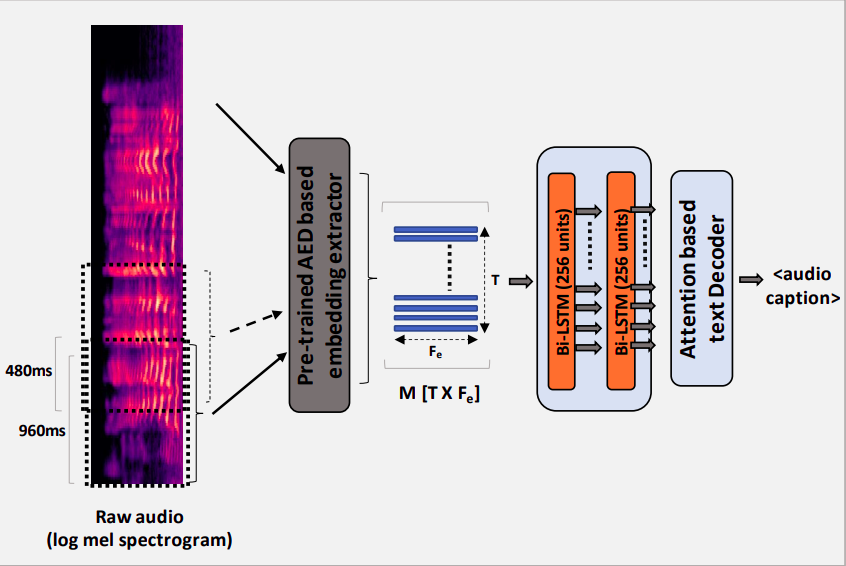}%}
	%  \vspace{2.0cm}
	\caption{Proposed AAC encoder-decoder architecture.}%\medskip
	\label{fig:enc_dec_model}
	\vspace{-0.5cm}
\end{figure}

\vspace{-0.3cm}
\label{sec:system_design}
The input to AAC task constitutes a temporal context and hence is modeled as a seq2seq problem where, a sequence of words, $W=[w_1, w_2, ..., w_n]$ ($n$ is the total number of words in the caption) is to be generated corresponding to a given sequence of audio samples, $A = [a_1, a_2, ..., a_f]$ ($f$ is the total number of frames in the audio). The encoder generates an encoding, $E$, from the input audio features, which is later used by the decoder along with the token at previous time-step. We first discuss the pre-trained AED encoders used in our work, followed by a detailed description of the AAC encoder and the attention based decoder (see Fig. \ref{fig:enc_dec_model}).
\subsection{Pre-trained Audio encoders}
Existing AAC datasets utilize crowd-sourcing platforms for obtaining captions, which makes the annotation task time-consuming, thus hinders the availability of large scale AAC datasets. Additionally, often the size of vocabulary for the existing AAC datasets makes it challenging to learn generic captions for audio samples outside the dataset. However, large scale datasets for Audio Event Detection (AED) tasks are available in the public domain, which has made the training of large AED models feasible. In order to overcome the small size of the AAC datasets, we adopt a transfer learning approach to incorporate the existing information regarding generic audio events from pre-trained AED models within the AAC training paradigm. We experiment with two such pre-trained models trained on the Audioset Dataset ($>10$x larger compared to existing AAC datasets), namely, YAMNet and Audio Spectrogram Transformer (AST). The process of extracting embeddings relevant to our task of AAC is detailed below.

\subsubsection{YAMNet \cite{yamnet2019}}
\begin{figure*}[t]
	\centering
	%\centerline{
	\includegraphics[width=0.9\textwidth]{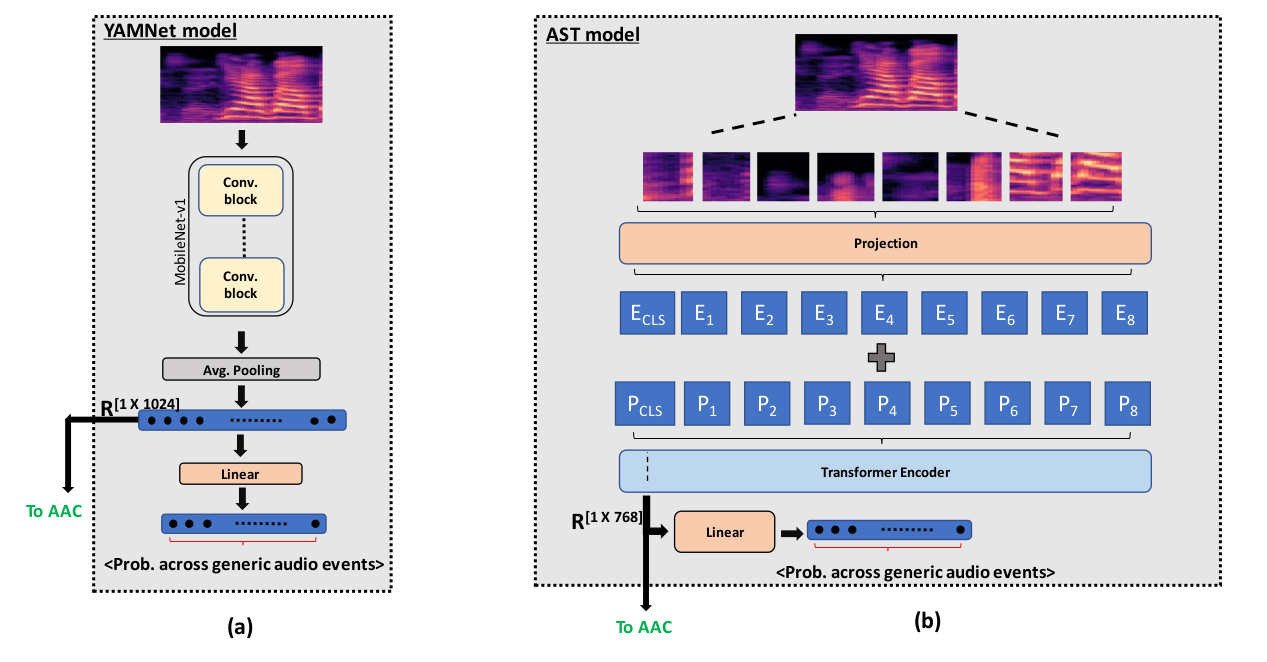}%}
	%  \vspace{2.0cm}
	\caption{Pre-trained AED based embedding extractors. (a) Penultimate layer output from the YAMNet model is used as the AED based embedding for the AAC task, (b) In the Audio Spectrogram Transformer (AST) model, transformer encoder's output corresponding to the \CLS token is used as AED based embedding. ($E_{i}$ and $P_i$ refer to the 1-D patch embedding and positional embedding respectively, as mentioned in Section \ref{sec:AST_details})}%\medskip
	\label{fig:aed_embs}
	\vspace{-0.5cm}
\end{figure*}

\vspace{-0.2cm}
YAMNet is a pre-trained AED model trained over 521 unique audio events from the Audioset ontology \cite{gemmeke2017audio}. YAMNet is based on MobileNet architecture which uses log Mel spectrograms corresponding to 960 msec of audio slices as input and outputs softmax probability scores over a set of generic audio events. Instead of the final output we use the intermediate embedding of 1024 dimension from the penultimate layer as the encoding for our downstream AAC task. (see Fig. \ref{fig:aed_embs}(a))

\subsubsection{Audio Spectrogram Transformer (AST) \cite{gong2021ast}}
\label{sec:AST_details}\vspace{-0.2cm}
AST is a convolution-free transformer based model for AED, trained upon the Vision Transformer (ViT) model for audio classification. We use the AST model pre-trained on the Audioset recipe. Unlike the YAMNet model, AST splits the input log Mel filterbank features into a sequence of 16
x 16 patches which are further flattened to form a 1D patch embedding and fed as input to the transformer encoder (see Fig. \ref{fig:aed_embs}(b)) along with a \CLS token appended at the beginning. The transformer encoder output corresponding to the \CLS token is used as the audio encoding for our downstream AAC task since it serves as the audio spectrogram representation \cite{gong2021ast}.

During training of the AAC (downstream task), we freeze the layers corresponding to the pre-trained AED models. The embedding (extracted from the intermediate layer of the AED models) for consecutive audio segments of size $w$, (with 50\% overlap) are concatenated along the time axis before providing as input to the AAC encoder (discussed in section \ref{subsec:aac_encoder}). This results in an matrix, $M$ of dimension $T \mbox{x} F_e$, where $T$ corresponds to the total number of consecutive segments in the input audio file, and $F_e$ denotes the feature embedding from pre-trained AED models (1024 in case of YAMNet and 768 in case of AST).

\subsection{AAC Encoder}
\label{subsec:aac_encoder}
\vspace{-0.2cm}
A generated caption is expected to capture the temporal relationships of the individual audio events such as, "\textbf{After} the lighter clicks and lights the cigarette the man coughs", "music is played \textbf{before} trailing off into silence". Preserving the temporal context in the AED based embeddings (the input to AAC encoder) is important since the input audio files among AAC tasks are often 15-20 sec long, and most audio events are not present throughout the duration of entire audio, but rather limited to specific segments. Hence, we refrain from averaging the AED based embedding, $M$, along the temporal axis in case of YAMNet or passing the entire audio as input to the AST transformer encoder to obtain a 1-D embedding corresponding to the entire audio file. Instead the AAC encoder consists of a dedicated recurrent layer that can learn the temporal contexts within the consecutive segments of audio. We use a stack of two Bi-directional LSTM layers with 256 units each. The outputs at each time-step from the second LSTM layer is used as encoded feature, and passed to the decoder.

\subsection{Attention based decoding}
\vspace{-0.2cm}
LSTM-RNN networks have been extensively explored in modeling sequential data points within various translation task, speech recognition, video captioning etc. A LSTM unit, comprises of a cell state (i.e. memory) and three gates, forget gate $f_t$, input gate $i_t$ and an output gate $o_t$, each with learnable weight matrices that help decide the information to be omitted from the previous step, the information to be incorporated into the cell state from the current unit's input and the information to be passed onto the next unit (hidden state), respectively. During the decoding process, the LSTM decoder generates each word step-by-step using the hidden state of the LSTM unit, and an attention weighted encoding $E_{attn}$, of the encoder's output $E$. The intuition behind $E_{attn}$ is that the decoder weighs different segments of the input audio separately, while generating each word. $E_{attn}$ is computed by multiplying the attention weights with the encoder output $E$, which are computed by applying a ReLU activation over the decoder's last output and $E$.
\vspace{-0.2cm}
\begin{equation}
\alpha = ReLU(EW_{e} + EW_{h})
\label{eq:alpha_eq}
\end{equation}
%\vspace{0.1cm}
\begin{equation}
%\[
A = Softmax(\alpha W_{a})
%\]
\end{equation}
\begin{equation}
%\[
E_{attn} = A * E
%\]
\end{equation}
%\[
%A = Softmax(\alpha W_{a})
%\]
%\[
%E_{attn} = A * E
%\]
where, 
$W_{e} \in \Re^{d_{e} \times d_{a}}$, $W_{h} \in \Re^{d_{h} \times d_{a}}$, $W_{a} \in \Re^{d_{a} \times 1}$ are the learnable weight matrices, and
$d_{e}, d_{a}, d_{h}$ are the latent dimensions for encoder vector, attention vectors, hidden state vector and encoder output $E$, respectively. At each decoding step, the encoder's output is weighed using learned attention weights based on the decoder's previous hidden state output, and finally generating a new word in the decoder with the previous word and the attention weighted encoding.

\section{Experiments}
\label{sec:experiments}
\vspace{-0.2cm}
To validate our proposed architecture we experiment using the already provided training, validation and evaluation splits for the Clotho dataset
\cite{drossos2020clotho}. 
We report our results in terms of standard metrics for natural language generation tasks, that have been widely accepted in the existing literature on the AAC task
as well.

\subsection{Dataset and Preprocessing}\vspace{-0.2cm}
Clotho \cite{drossos2020clotho} is an audio captioning dataset consisting of audio files from the Freesound \cite{fonseca2017freesound} platform. Unlike other AAC datasets, like Audiocaps \cite{kim2019audiocaps}, the annotation process in Clotho caters to the perceptual ambiguity w.r.t. to audio events since the annotation was performed using only acoustic cues with no context information (ex: visual cues)
provided to the annotators  \cite{drossos2020clotho}. Additionally, unlike the older datasets with only single captions for each audio sample, Clotho enables learning the diversity in captions by providing 5 crowd-sourced captions for each audio file, each ranging between 8-20 words in length. To maintain consistency with previous work, we use the same splits provided as the part of Detection and Classification of Acoustic Scenes and Events (DCASE) 2020 challenge \cite{DCASE2020}, which consists of 2893 development samples and 1043 evaluation samples, each ranging between 15-30 sec in duration. All the audio files are down-sampled to 16 kHz sampling frequency, to be compatible with the pre-trained AED models. We adopt a bucket padding technique, wherein the audio samples are padded to match the longest duration within the training batch only. We compute the vocabulary for our task by choosing the words with at least 10 occurrences among all the captions in the dataset. A \START and \END token is added to the beginning and end of each caption respectively, and words that are not a part of the vocabulary are replaced using a \UNK token. We also append \PAD tokens such that each caption has maximum 20 tokens.
\vspace{-0.4cm}
\subsection{Training}
\vspace{-0.2cm}
During training, SpecAugment is used with max time mask length of 192 frames and max frequency mask length of 48 bins with 0.4 probability. The model is trained using a batch size of 32 with categorical cross-entropy computed at each decoding step between the probability distribution over the entire vocabulary and the one-hot encoded word vector corresponding to the actual word in that step. We use Adam optimizer with initial learning rate of 1e-4 which is reduced by 50\% when 3 consecutive epochs with no change in the validation BLEU scores are encountered.
  
\begin{table*}[]
	
	%\centering
	\tiny
	\caption{Performance of the proposed approach and previous methods on the Clotho evaluation set. For all metrics, a higher score reflects
better performance. (B-N: N-gram BLEU score.)}
	\vspace{0.05cm}
	\label{table:Results}
	%	\begin{adjustbox}{max width=\linewidth}
	\resizebox{\linewidth}{!}{
		\begin{tabular}{|l|l|l|l|l|l|l|l|l|l|}
			\hline
			\textbf{Model} & \textbf{Augment} & \textbf{B-1} & \textbf{B-2} & \textbf{B-3} & \textbf{B-4} & \textbf{ROUGE-L} & \textbf{CIDEr} & \textbf{METEOR} &\textbf{SPICE}\\ \hline
			Koizumi et. al. \cite{koizumi2020transformer} & - & 52.1 & 30.9 & 18.8 & 10.7 & 34.2 & 25.8 & 14.9 & 9.7 \\ \hline
			Xu et. al.\cite{xu2021investigating} & - & 55.6 & 36.3 & 24.2 & 15.9 & 36.8 & \textbf{37.7} & 16.9 & \textbf{11.5} \\ \hline \hline
			Baseline - Log mel spectrogram & SpecAugment & 48.2 & 27.7 & 16.4 & 11.2 & 31.8 & 20.7 & 13.8 & 8.8 \\ \hline
			YAMNet $F_e$ + Attn. decoder &  - & 53.6 & 35.3 & 21.8 & 14.7 & 34.1 & 33.5 & 15.1 & 10.3 \\ \hline
			YAMNet $F_e$ + Attn. decoder & SpecAugment & 54.8 & 35.8 & 23.4 & 15.2 & 34.3 & 34.7 & 15.7 & 10.7 \\ \hline
			AST $F_e$ + Attn. decoder &  - & 55.3 & 36.1 & \textbf{24.4} & 15.7 & 37.1 & 36.6 & 16.4 & 10.9\\ \hline
			AST $F_e$ + Attn. decoder & SpecAugment & \textbf{55.8} & \textbf{36.8} & 24.3 & \textbf{16.1} & \textbf{37.3} & 37.1 & \textbf{17.3} & \textbf{11.3}\\ \hline
	\end{tabular}}
	\vspace{-0.5cm}
\end{table*}
\subsection{Evaluation}
\vspace{-0.2cm}
\begin{figure}[h]
	%\vspace{-0.4cm}
	\centering
	%\centerline{
	\includegraphics[width=0.5\textwidth]{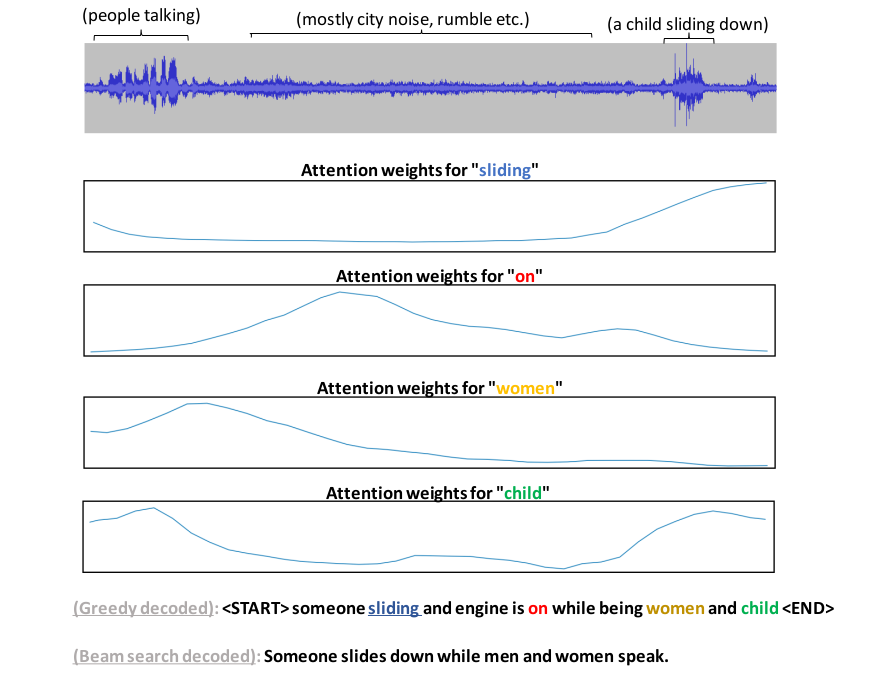}%}
	%  \vspace{2.0cm}
	\caption{Audio file: {\tt evaluation/Downtown Montreal.wav}. The top subplot depicts the raw audio, with events localized based on our perception (for reference only). In all attention sub-plots above, the x-axis resembles the number of time frames in the AED based encoding, $M$ and y-axis represents the value of $\alpha$.}% for words inside " ".}%\medskip
	\label{fig:analysis}
	\vspace{-0.5cm}
\end{figure}
In an extensive experimental evaluation, we adopt the metrics used in the DCASE 2020 Challenge, i.e., BLEU, ROUGE-L, METEOR, CIDEr, SPICE. Among all the metrics, BLEU (geometric mean of the modified n-gram score), although fast, is the least correlated to human perception. ROUGE-L measures the precision and recall using the longest matching sequence of words. CIDEr assumes the words occurring frequently (calculated using tf-idf, term frequency-inverse document frequency) in captions are less informative and hence weighs them lesser during reference matching. METEOR computes the F-score using exact matches, matches after applying the stemming operation and matches within synonyms. For obtaining the final captions, instead of greedy decoding (choosing the word with highest probability at every decoding step), the output probability distribution over the entire vocabulary for each decoding step is passed as an input to a beam search decoder with beam size of 3.

We compare our approaches using YAMNet and AST based AED embedding extractors, while keeping the AAC encoder-decoder architecture same with already existing approaches. 
We use the same architecture but trained from scratch using log mel-spectrogram features as inputs to the AAC encoder as our baseline model. From table \ref{table:Results} it can be seen that our model with AST based embedding extractor surpasses existing approaches in majority of the 
%translation 
metrics. Between the YAMNet and AST based models, the latter outperforms by a considerable margin, owing to its better performance on the AED tasks. Also, when comparing same models trained with and without augmentation, the advantage offered by SpecAugment to regularize models on the small AAC dataset (Clotho: ~3000 training samples) is clearly visible. We hypothesize that the lesser CIDEr score can be attributed to the smaller size of vocabulary chosen during training, thereby forcing our model to predict \UNK tag instead of words that are less frequent. A larger vocabulary means lesser \UNK tags are used during the training, however, it increases the size of the word embedding matrix inside the decoder and vice-versa.

\subsection{Analyzing temporal attention weights}
\vspace{-0.2cm}
To testify temporal attention during the decoding step, we visualize the attention weights, $\alpha$ (\ref{eq:alpha_eq}), that are computed across the total number of frames in $M$, while generating each token. It is important to note
that the overall captioning performance depends on 
(a) the ability of our model to localize unique audio events, and 
(b) the ability to model language.
% modeling capability. 
The intuition behind temporal attention is to guide the decoding step by weighing sections within entire audio
%non-uniformly 
based on the audio events they represent. 
Hence, as a part of the analysis (see Fig. \ref{fig:analysis}) we use greedy
decoding (selecting only the word with maximum probability at each decoding
step) instead of the beam search decoder to eliminate the contribution of language model
during captioning. 
In Fig. \ref{fig:analysis}, we plot the attention weights obtained for specific keywords 
(sliding, on, women, child) 
generated as a part of the decoded caption. The 
%non-uniform distribution of the 
change in weights along the temporal axis suggests that the model inherently 
uses sections of the audio file as cues to generate keywords related to the captured audio events. It
can be also observed from the attention plot of the keyword "sliding" 
%From the first attention subplot it is clear 
that, though the word "sliding" (in caption) is not aligned with the sequence of occurrence of audio events (audio
event "people talking" happens before the audio event "sliding") in the actual audio file, the prediction of the word "sliding" is still attributed to the latter part of the actual audio.  

\section{Conclusion}
\label{sec:conclusion}
\vspace{-0.3cm}
In this paper, we showcase that an encoder-decoder architecture with only Bi-LSTM layers when combined with temporal attention and augmentation can yield improvements over existing computationally intensive architectures %with higher learnable parameters 
when evaluated over standard translation metrics. We also compared the use of two existing AED models as pre-trained embedding extractors for the task of caption generation. Our experiments show that both the variants surpass our baseline model trained from scratch using only log mel-spectrogram features by a huge margin. Similar to any audio translation task, caption generation performance depends on the ability of the acoustic model to localize unique audio events, and the language modeling aspect of the decoder to form meaningful sentences using the detected events as keywords. Although, the latter has been previously shown attainable using beam search decoders for learning the language context information, we validate the former attribute for our model, by analyzing the attention weights generated across input time frames in our AED based embedding when generating each token in the caption.

%\newpage
\bibliographystyle{IEEEbib}
\bibliography{refs}

\begin{thebibliography}{10}

\bibitem{drossos2017automated}
Konstantinos Drossos, Sharath Adavanne, and Tuomas Virtanen,
\newblock ``Automated audio captioning with recurrent neural networks,''
\newblock in {\em IEEE Workshop on Applications of Signal Processing to Audio
  and Acoustics (WASPAA) 2017}, pp. 374--378.

\bibitem{oncescu2021audio}
Andreea-Maria Oncescu, A~Koepke, Jo{\~a}o~F Henriques, Zeynep Akata, and Samuel
  Albanie,
\newblock ``Audio retrieval with natural language queries,''
\newblock {\em arXiv preprint arXiv:2105.02192}, 2021.

\bibitem{wu2019enhancing}
Yuzhong Wu and Tan Lee,
\newblock ``Enhancing sound texture in cnn-based acoustic scene
  classification,''
\newblock in {\em IEEE ICASSP 2019)}, pp. 815--819.

\bibitem{lu2015context}
Tong Lu, Gongyou Wang, and Feng Su,
\newblock ``Context-based environmental audio event recognition for scene
  understanding,''
\newblock {\em Multimedia Systems}, 2015.

\bibitem{hong2010dynamic}
Richang Hong, Meng Wang, Mengdi Xu, Shuicheng Yan, and Tat-Seng Chua,
\newblock ``Dynamic captioning: video accessibility enhancement for hearing
  impairment,''
\newblock in {\em Proceedings of the ACM international conference on Multimedia
  2018}, pp. 421--430.

\bibitem{portelo2009non}
Jose Portelo, Miguel Bugalho, Isabel Trancoso, Joao Neto, Alberto Abad, and
  Antonio Serralheiro,
\newblock ``Non-speech audio event detection,''
\newblock in {\em IEEE ICASSP 2009}, pp. 1973--1976.

\bibitem{babaee2017overview}
Elham Babaee, Nor~Badrul Anuar, Ainuddin~Wahid Abdul~Wahab, Shahaboddin
  Shamshirband, and Anthony~T Chronopoulos,
\newblock ``An overview of audio event detection methods from feature
  extraction to classification,''
\newblock {\em Applied Artificial Intelligence}, vol. 31, no. 9-10, pp.
  661--714, 2017.

\bibitem{drossos2020clotho}
Konstantinos Drossos, Samuel Lipping, and Tuomas Virtanen,
\newblock ``Clotho: An audio captioning dataset,''
\newblock in {\em IEEE ICASSP 2020}, pp. 736--740.

\bibitem{kim2019audiocaps}
Chris~Dongjoo Kim, Byeongchang Kim, Hyunmin Lee, and Gunhee Kim,
\newblock ``Audiocaps: Generating captions for audios in the wild,''
\newblock in {\em Proceedings of the 2019 Conference of the North American
  Chapter of the Association for Computational Linguistics: Human Language
  Technologies, Volume 1 (Long and Short Papers)}, pp. 119--132.

\bibitem{chen2019temporal}
Jingwen Chen, Yingwei Pan, Yehao Li, Ting Yao, Hongyang Chao, and Tao Mei,
\newblock ``Temporal deformable convolutional encoder-decoder networks for
  video captioning,''
\newblock in {\em Proceedings of the AAAI conference on artificial
  intelligence}, 2019, vol.~33, pp. 8167--8174.

\bibitem{gregor2014deep}
Karol Gregor, Ivo Danihelka, Andriy Mnih, Charles Blundell, and Daan Wierstra,
\newblock ``Deep autoregressive networks,''
\newblock in {\em ICML 2014}. PMLR, 2014, pp. 1242--1250.

\bibitem{xu2021investigating}
Xuenan Xu, Heinrich Dinkel, Mengyue Wu, Zeyu Xie, and Kai Yu,
\newblock ``Investigating local and global information for automated audio
  captioning with transfer learning,''
\newblock in {\em IEEE ICASSP 2021}, pp. 905--909.

\bibitem{koizumi2020audio}
Yuma Koizumi, Yasunori Ohishi, Daisuke Niizumi, Daiki Takeuchi, and Masahiro
  Yasuda,
\newblock ``{Audio Captioning using Pre-Trained Large-Scale Language Model
  Guided by Audio-based Similar Caption Retrieval},''
\newblock {\em arXiv preprint arXiv:2012.07331}, 2020.

\bibitem{mei2021encoder}
Xinhao Mei, Qiushi Huang, Xubo Liu, Gengyun Chen, Jingqian Wu, Yusong Wu,
  Jinzheng Zhao, Shengchen Li, Tom Ko, H~Lilian Tang, et~al.,
\newblock ``An encoder-decoder based audio captioning system with transfer and
  reinforcement learning,''
\newblock {\em arXiv preprint arXiv:2108.02752}, 2021.

\bibitem{fonseca2018general}
Eduardo Fonseca, Manoj Plakal, Frederic Font, Daniel~PW Ellis, Xavier Favory,
  Jordi Pons, and Xavier Serra,
\newblock ``General-purpose tagging of freesound audio with audioset labels:
  Task description, dataset, and baseline,''
\newblock {\em arXiv preprint arXiv:1807.09902}, 2018.

\bibitem{barchiesi2015acoustic}
Daniele Barchiesi, Dimitrios Giannoulis, Dan Stowell, and Mark~D Plumbley,
\newblock ``Acoustic scene classification: Classifying environments from the
  sounds they produce,''
\newblock {\em IEEE Signal Processing Magazine}, vol. 32, no. 3, pp. 16--34,
  2015.

\bibitem{koizumi2020transformer}
Yuma Koizumi, Ryo Masumura, Kyosuke Nishida, Masahiro Yasuda, and Shoichiro
  Saito,
\newblock ``A transformer-based audio captioning model with keyword
  estimation,''
\newblock {\em Interspeech 2020}.

\bibitem{ozkaya2021audio}
Ay{\c{s}}eg{\"u}l {\"O}zkaya~Eren and Mustafa Sert,
\newblock ``{Audio Captioning with Composition of Acoustic and Semantic
  Information},''
\newblock {\em International Journal of Semantic Computing}, vol. 15, no. 02,
  pp. 143--160, 2021.

\bibitem{hossain2019comprehensive}
MD~Zakir Hossain, Ferdous Sohel, Mohd~Fairuz Shiratuddin, and Hamid Laga,
\newblock ``A comprehensive survey of deep learning for image captioning,''
\newblock {\em ACM Computing Surveys (CsUR)}, vol. 51, no. 6, pp. 1--36, 2019.

\bibitem{xu2015show}
Kelvin Xu, Jimmy Ba, Ryan Kiros, Kyunghyun Cho, Aaron Courville, Ruslan
  Salakhudinov, Rich Zemel, and Yoshua Bengio,
\newblock ``Show, attend and tell: Neural image caption generation with visual
  attention,''
\newblock in {\em ICML 2015}, pp. 2048--2057.

\bibitem{yamnet2019}
\url{https://github.com/tensorflow/models/tree/master/research/audioset/yamnet}
  Manoj~Plakal, Dan~Ellis,
\newblock ``{YAMNet},''
\newblock {\em GitHub repository}.

\bibitem{gong2021ast}
Yuan Gong, Yu-An Chung, and James Glass,
\newblock ``Ast: Audio spectrogram transformer,''
\newblock {\em arXiv preprint arXiv:2104.01778}, 2021.

\bibitem{gemmeke2017audio}
Jort~F Gemmeke, Daniel~PW Ellis, Dylan Freedman, Aren Jansen, Wade Lawrence,
  R~Channing Moore, Manoj Plakal, and Marvin Ritter,
\newblock ``Audio set: An ontology and human-labeled dataset for audio
  events,''
\newblock in {\em IEEE ICASSP 2017}, pp. 776--780.

\bibitem{fonseca2017freesound}
Eduardo Fonseca, Jordi Pons~Puig, Xavier Favory, Frederic Font~Corbera, Dmitry
  Bogdanov, Andres Ferraro, Sergio Oramas, Alastair Porter, and Xavier Serra,
\newblock ``Freesound datasets: a platform for the creation of open audio
  datasets,''
\newblock in {\em Hu X, Cunningham SJ, Turnbull D, Duan Z, editors. Proceedings
  of International Society for Music Information Retrieval (ISMIR) 2017}.

\bibitem{DCASE2020}
\url{http://dcase.community/challenge2020/task-automatic-audio-captioning},
\newblock ``{DCASE 2020 Challenge Task 6: Automated Audio Captioning,},''
\newblock .

\end{thebibliography}
\end{document}